\documentclass[pra,twocolumn,superscriptaddress]{revtex4}%
\usepackage{xcolor}
\usepackage{amsfonts,amsmath,amsthm}
\usepackage{bm}
\usepackage{mathrsfs}
\usepackage{dsfont}
\usepackage{graphicx}
\usepackage{hyperref}
\usepackage{epstopdf}
\usepackage{amsfonts}
\usepackage{pifont,units}
\usepackage{algorithm,algorithmicx,algpseudocode}
\usepackage{braket}
\usepackage{hyperref}
\usepackage[qm]{qcircuit}
\newtheorem{problem}{Problem}

\newtheorem{theorem}{Theorem}
\newtheorem{lemma}[theorem]{Lemma}

\begin{document}
\title{Quadratic Quantum Speedup for  Perceptron Training}
\author{Pengcheng Liao}
\affiliation{Institute for Quantum Science and Technology, University of Calgary, Alberta, T2N 1N4, Canada }

\author{Barry C. Sanders}
\email{sandersb@ucalgary.ca}
\affiliation{Institute for Quantum Science and Technology, University of Calgary, Alberta, T2N 1N4, Canada }
\affiliation{%
Shanghai Branch, National Laboratory for Physical Sciences at Microscale,
University of Science and Technology of China, Shanghai 201315, China%
}
\affiliation{%
CAS Center for Excellence and Synergetic Innovation Center in Quantum Information and Quantum Physics, University of Science and Technology of China, Hefei, Anhui 230026, China%
}
\author{Tim Byrnes}
\email{tim.byrnes@nyu.edu}
\affiliation{New York University Shanghai, 1555 Century Ave, Pudong, Shanghai 200122, China}  
\affiliation{State Key Laboratory of Precision Spectroscopy, School of Physical and Material Sciences, East China Normal University, Shanghai 200062, China}
\affiliation{NYU-ECNU Institute of Physics at NYU Shanghai, 3663 Zhongshan Road North, Shanghai 200062, China}
\affiliation{National Institute of Informatics, 2-1-2 Hitotsubashi, Chiyoda-ku, Tokyo 101-8430, Japan}
\affiliation{Department of Physics, New York University, New York, NY 10003, USA}

\begin{abstract}
Perceptrons, which perform binary classification, are the fundamental building blocks of neural networks. 
Given a data set of size~$N$ and margin~$\gamma$ (how well the given data are separated), the query complexity of the best known quantum training algorithm scales as either $(\nicefrac{\sqrt{N}}{\gamma^2})\log(\nicefrac1{\gamma^2)}$ or $\nicefrac{N}{\sqrt{\gamma}}$, which is achieved by a hybrid of classical and quantum search.
In this paper, we improve the version space quantum training method for perceptrons such that the query complexity of our algorithm scales as $\sqrt{\nicefrac{N}{\gamma}}$.
This is achieved by constructing an oracle for the perceptrons using quantum counting of the number of data elements that are correctly classified.  We show that query complexity to construct such an oracle has a quadratic improvement over classical methods. Once such an oracle is constructed, bounded-error quantum search can be used to search over the hyperplane instances.   The optimality of our algorithm  is proven by reducing the evaluation of a two-level AND-OR tree (for which the query complexity lower bound is known) to a multi-criterion search.
Our quantum training algorithm can be generalized to train more complex machine learning models such as neural networks,
which are built on a large number of perceptrons.
\end{abstract}

\date{\today}
\maketitle

\section{Introduction}

Quantum computing has been shown to have an algorithmic speedup~\cite{grover1996fast,shor1994factor, Harrow2009Quantum}
in comparison to classical computers for particular tasks.
In particular, quantum unsorted database searching~\cite{grover1996fast} has been proven to be quadratically faster in terms of query complexity than the best possible classical algorithm.
Shor's algorithm~\cite{shor1994factor} is exponentially faster compared to the known best classical algorithm. 
Meanwhile, machine learning is an important and powerful tool in computer science for pattern-searching.  
More recently, there has been intense investigation of quantum machine learning (QML):
designing machine learning algorithms tailored for quantum computers, aiming to construct methods that have the advantage of both quantum superposition and machine learning
\cite{2014Lloyd_qprinciple,rebentrost2014quantum,lloyd2013quantum,wiebe2018quantum,biamonte2017quantum,dunjko2016quantum,lloyd2018quantum,schuld2019quantum,amin2018quantum}. 

\begin{figure}
	\centering
	\includegraphics[width=1.0\linewidth]{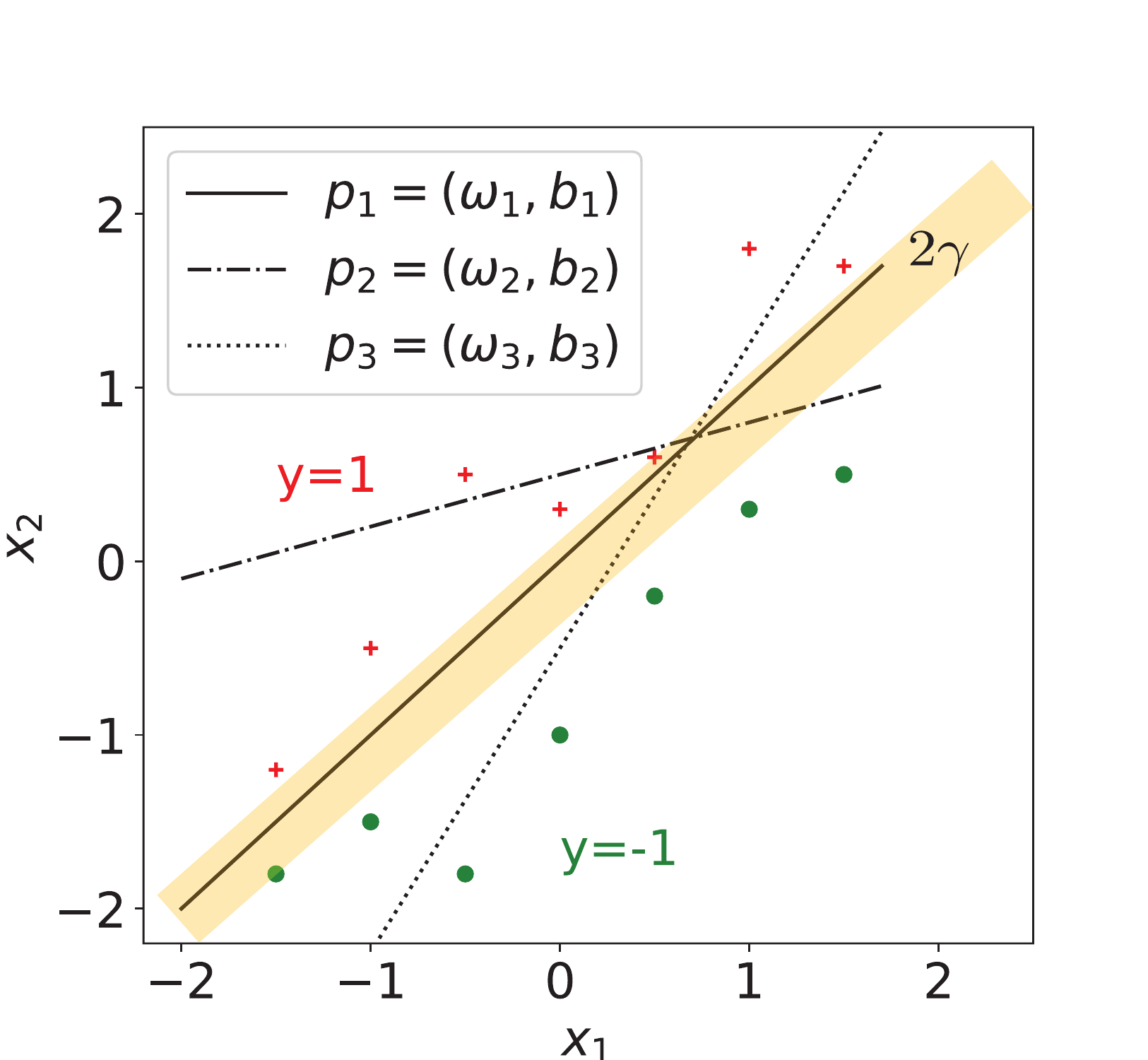}
	\caption{A two-dimensional example of perceptron classification.  
Two classes of data are shown: data with $y=+1$ are labeled by $+$ and  data with $y=-1$ are labeled by a solid circle.  In this example, the line $p_1$ separates the data perfectly while $p_2$ and $p_3$ do not. This corresponds to a $N=12,K=3,M=2, 2\gamma = 0.39$ case in Problem~\ref{pro:perceptron_version-space_training}. }
	\label{fig:perceptron}
\end{figure}

As the building block of many more advanced machine learning techniques~\cite{Haralick1973Transactions,friedman2001elements,tezak2015coherent}, 
the perceptron is a simple supervised machine learning model~\cite{rosenblatt1958perceptron, minsky1969introduction}, which acts as a binary classifier for multi-dimensional labeled data.
A two-dimensional (2D) example of a perceptron classification is given in Fig.~\ref{fig:perceptron}.  The aim of perceptron training is to find a hyperplane (for the 2D case a line) such that  all data are separated perfectly into two groups.  
Realistic applications include converting images/audio/video to multi-dimensional real vectors with labels. Once trained, perceptrons can be used to classify new data into their respective categories.
A commonly used classical training algorithm is online training (the data are processed one by one): first choose an arbitrary hyperplane and check in sequence whether it classifies any data point incorrectly; if such a data point is found, then update the hyperplane accordingly; repeat this procedure until all data are classified correctly. 
The online learning algorithm is guaranteed to converge to an answer if the training data set is linearly separable.

There have been several proposals investigating how  perceptrons can be constructed in the quantum regime~\cite{tacchino2019artificial,schuld2015simulating,du2018implementable,torrontegui2019unitary,behrman2000simulation,cao2017quantum,wiersema2019implementing}.
Some of the proposed quantum models still use the online training algorithm~\cite{schuld2015simulating,behrman2000simulation}.
For example, replacing a subroutine in online training with amplitude amplification can reduce the query complexity from $O\left(\frac{N}{\gamma^2}\log\frac1{\gamma^2}\right)$ to $O\left(\frac{\sqrt{N}}{\gamma^2}\log\frac1{\gamma^2}\right)$~\cite{kapoor2016quantum}, where~$N$ is the size of the data set and~$\gamma$ is the margin.
The margin~$\gamma$ of a data set  quantifies how far the labeled two classes are separated.
The step-by-step regression used in online training is highly classical and does not fully take advantage of quantum parallelism.
Another approach to training perceptrons is based on the version space interpretation~\cite{kapoor2016quantum}.
In version space training, a certain number of hyperplanes is sampled and then checked in sequence to see if any of the samples are in the version space.  This can be considered as a search task: find the hyperplane that is in the version space among the sampled ones.
For a data set of size~$N$ and  margin~$\gamma$, the query complexity of classical version space training is  $O\left(\nicefrac{N}{\gamma}\right)$, which 
can be reduced to $O(\nicefrac{N}{\sqrt{\gamma}})$ by replacing a classical subroutine with amplitude amplification ~\cite{kapoor2016quantum}; thus, achieving a quadratic speedup with respect to the inverse of the margin.
However, the query complexity still scales linearly with the size of the training data set $N$, which is typically a large number. 
It therefore would be desirable to have a quantum training algorithm with a better scaling of query complexity with respect to $N$.

In this paper, we propose a quantum algorithm for which the query complexity is $\Theta\left(\sqrt{\nicefrac{N}{\gamma}}\right)$. The main task in our approach will be to construct an oracle $ U_g $ that returns a Grover phase~$-1$ for hyperplanes that are in the version space, and $+1$ otherwise. Such an oracle yields the $\nicefrac1{\sqrt{\gamma}}$ scaling, but previously has not been shown to be constructible with $  \sqrt{N}$ scaling. To do this, we use an algorithm closely related to quantum counting, where it is possible to find the number of solutions to a search task. The counting is performed with respect how many data points are classified correctly for a given hyperplane.  The oracle $ U_g $ is then constructed in such a way that a phase~$-1$ is returned only when all the data points are classified correctly. We also show the optimality of our algorithm by showing the equivalency of version space training to a search task with more than one criterion, i.e., a multi-criterion search. This can then in turn be related to a two-level AND-OR tree \cite{Ambainis2010ANDOR},  for which the query complexity lower bound  is known~\cite{ambainis2002quantum}.  

Our paper is organized as follows: 
Sec.~\ref{sec:background}  reviews classical version space training and  quantum search on bounded-error input;
Sec.~\ref{sec:basicidea} gives the basic idea of our quantum version space training protocol;
Sec.~\ref{sec:qversion_training} describes in more detail our training algorithm and provides the relevant proofs; 
Sec.~\ref{sec:optimality} shows the optimality of our algorithm; and
Sec.~\ref{sec:conclusion} summarizes our results.

\section{Background}
\label{sec:background}

We now briefly explain the essential background related to the problem which we wish to solve, and the key techniques that we will employ.

\subsection{Perceptrons}
\label{sec:perceptron}

Consider a set of $ N $ data elements  $\{(\bm{x}_i,y_i)\} $, specified by an $ M$-dimensional real vector $ \bm{x}_i \in \mathbb{R}^M $ and its category  $y_i=\pm 1$.  A perceptron then  takes  $\bm{x} $ as its input and outputs a value $y=\pm 1$ according to the rule
\begin{equation}
y = \operatorname{sgn} (\bm{w} \cdot \bm{x}+b) = 
\begin{cases}
1,\quad& {\rm if} \quad\bm{w} \cdot \bm{x} +b\ge 0\\
-1,\quad& {\rm otherwise}.
\end{cases}
\label{perceptrondef}
\end{equation} 
The perceptron is parameterized by a hyperplane $\bm{p}=(\bm{w},b)\in\mathbb{R}^{M+1}$, consisting of a weight vector $\bm{w}\in\mathbb{R}^M$ and a bias $b\in\mathbb{R}$.

The data set  $\{(\bm{x}_i,y_i)\} $ is said to be separated by margin~$\gamma$ if there exists a hyperplane $(\bm{w},b)$ such that $|\bm{w}\cdot\bm{x}_i +b|\ge \gamma~$ holds for all $i $. The margin can be understood to quantify how well-separated the data are (Fig. \ref{fig:perceptron}).  
The margin is always a positive quantity $ \gamma > 0 $ for a linearly separable data set. 

The aim of perceptron training is then as follows.  Given~$N$ training data elements  $\{(\bm{x}_i,y_i)\}$ separated by margin $\gamma>0$, we wish to find a vector $\bm{p}=(\bm{w},b)$ such that 
\begin{align}
\left(\bm{w}\cdot \bm{x_i}+b\right)y_i>0  , 
\label{perceptcrit}
\end{align}
holds for all $i $.  With this condition satisfied, all data elements give the correct classification $ y_i $ when $ \bm{x}_i $ is substituted into Eq.~(\ref{perceptrondef}).

\subsection{Version Space Training}
\label{sec:classical_version_spa_tra}

Perceptron training can be reformulated as an equivalent search problem using version space training.  \textit{Version space} is defined as the set of hyperplanes $ \bm{p} $ that satisfy (\ref{perceptcrit}).  In version space training, one samples $K$ hyperplanes $\bm{p}_1,\ldots,\bm{p}_K$ and then checks if any~$\bm{p}_j$ is in the
version space. It was shown in Ref.~ \cite{kapoor2016quantum} that if the hyperplanes are drawn uniformly from a spherical Gaussian distribution and $K\in\Theta\left(\nicefrac{\log(1/\epsilon)}{\gamma}\right)$,
then there exists at least one hyperplane in the version space  with failure probability at most $\epsilon$.

Let us introduce some common notation that will be used throughout this paper.
The symbols $N, K\in\mathbb{N}$ are used to denote the size of the training data and hyperplanes, respectively,
and
\begin{equation}
    i\in[N]:=\Set{0,1,\ldots,N-1},\;
    j\in[K]
\end{equation}
are used to specify an instance of the training data and a hyperplane, respectively.  
For simplicity,
we assume $n = \log_2N$ and $k=\log_2K$ are integers unless otherwise stated. 
With $K$ sampled hyperplanes, perceptron training effectively becomes the following search problem.

\begin{problem}
\label{pro:perceptron_version-space_training}
Given~$N$ training data elements $\{(\bm{x}_i,y_i)\}$ and $K$ hyperplanes $\Set{\bm{p}_j}$, either output an index $j\in[K]$ such that~$\bm{p}_j$ is in the version space with bounded error or output~$-1$ if no such $j$ exists with bounded error. 
\end{problem}
Here we require the training problem to be solved with bounded error, i.e., the result is correct with probability greater or equal to $2/3$.

In order to solve this problem, let us assume that we have access to  a  classical oracle 
\begin{equation}
\label{oracle_f}
f(i,j)=
\begin{cases}
1, &\quad \bm{p}_j \quad \text{classifies}\quad \bm{x}_i\quad \text{correctly} \\
0, &\quad \bm{p}_j \quad \text{classifies}\quad \bm{x}_i\quad \text{incorrectly}
\end{cases}
\end{equation}
and its quantum counterpart
\begin{equation}
\label{eq:quantum_oracle}
U_f\Ket{i,j} = (-1)^{f(i,j)} \Ket{i,j}.
\end{equation} 
All algorithms solving Problem~\ref{pro:perceptron_version-space_training} will be  evaluated by how many times $f$ or $U_f$ is used, which is referred as the query complexity. 

Classically,  one can use $f$ to check $\Set{\bm{p}_j}$ in sequence until the answer is found, which requires $O(NK)$ queries to $f$.
The quantum algorithm  proposed in Ref.~\cite{kapoor2016quantum} made use of the following unitary
\begin{equation}
\label{eq:U_g}
U_g \Ket{j} =(-1)^{g(j)}\Ket{j}, 
\end{equation}
where $g(j)=\operatorname{AND}_f(j)=\bigwedge_{i\in[N]}f(i,j)$.
By calling (\ref{eq:quantum_oracle})~$N$ times, it is easy to construct $U_g$ in a classical way.
One then  uses the gate $U_g$ to perform quantum search to find the $j$ that indexes a vector in the version space, costing $O(N\sqrt{K})$ queries to $U_f$ in total.
In this paper, we adopt a different approach to construct $U_g$ with only $O(\sqrt{N})$ queries to $U_f$ by using quantum counting.

\subsection{Quantum search on bounded-error input}
\label{sec:qsearch&phasest}

We now briefly describe the bounded-error version of the quantum search algorithm, which will be used in our perceptron training algorithm described later.  

First we define the standard quantum search algorithm, i.e. Grover's algorithm.  Given a Boolean function $h:[N]\to\set{0,1}$ and quantum oracle $U_h$ such that 
\begin{align}
U_h\Ket{i}=(-1)^{h(i)}\Ket{i} , 
\end{align}
quantum search is asked to return a variable $i\in[N]$ such that $h(i)=1$ or return~$-1$ if no solution exists. 
If a solution to the search problem is promised to exist, the number of queries needed to find such an answer with bounded error is $\Theta(\sqrt{N/L})$, where $L=|h^{-1}(1)|$ is the number of solutions~\cite{grover1996fast,Boyer1998tight}.

The quantum search algorithm was later improved and generalized in various ways \cite{Giri2017}, and one of improvements is quantum search with bounded-error input \cite{Hoyer2003Quantum}.
In bounded-error quantum search, one considers an oracle $ U_{\tilde{h}} $ where there is a probability of failiure.  It was shown that the search problem can still be solved with the same query complexity with the quantum oracle modified to
\begin{align}
U_{\tilde{h}} \Ket{i}= (-1)^{\tilde{h}(i)}\Ket{i} 
\label{imperfectoracle}
\end{align}
where $ \tilde{h}(i) = h(i) $ with probability $ \ge 2/3 $ for all $ i $.  By interleaving the amplitude amplification and error reduction and using $O(\sqrt{N})$ queries of (\ref{imperfectoracle}), one can obtain either a variable $i$ evaluated to be 1 with bounded error or~$-1$ when there is no such variable .
For further details, we refer readers to Ref.~\cite{Hoyer2003Quantum}.
The quantum search algorithm on bounded-error input in the rest of this paper is referred as
\begin{equation}
    \textsc{bEqseArch}(n,U_{\tilde{h}}) \to i \in[N]~\text{or} -1 .
\end{equation}
%

\section{Quantum version space training: the basic idea}
\label{sec:basicidea}

Before presenting the more formal version of our algorithm, we first present a simplified version of the argument which shows the basic idea of our approach.  

To get some intuition, let us first consider an elementary version space training problem with $ N = 4 $ data elements and $ K = 3$ hyperplanes. The function $ f $ is, for example, 
\begin{equation}
\label{eq:examplesize4}
f =
\begin{bmatrix}
0 & 1 & 1 \\
 0 & 1 & 1 \\
0 & 0 & 1 \\
 1 & 1 & 1 
\end{bmatrix}.
\end{equation}
The function values are listed in the matrix, where the $i$ labels the data elements in the rows and the $j$ labels the hyperplanes in the column.  For the example above, $ j = 3$ is a solution of the version space problem, because all the data elements $ i $ are classified correctly. Our aim is to construct the $ g $ function, which for this example is
\begin{equation}
g =
\begin{bmatrix}
0 & 0 & 1
\end{bmatrix} .
\end{equation}
Clearly, this can be constructed by taking the product of all the elements in a column of $ f $.  Once the $ g $ function is constructed, a quantum search can find the solution with $  \propto \sqrt{K} $ queries.  The main problem is how to construct $ U_g $ in such a way that it improves upon the previously obtained result with $ \propto N  $ queries.  
The key idea of our proposal is to use quantum counting to evaluate the product of each column instead of checking every entry in the column one by one, which gives a quadratic speedup in the scaling of $N$. 

We now give the basic argument for how to construct $U_g$ using only  $O(\sqrt{N})$ queries of $U_f$. Consider the initial state which is an equal superposition state of all data elements $ i $ and choose a particular hyperplane $ j $ (i.e., one of the columns of (\ref{eq:examplesize4})).  This state can be rewritten
\begin{equation}
\label{eq:sincosstate}
\Ket{+}^{\otimes n}\Ket{j} = \sin \theta_j  \Ket{g_j }\Ket{j} + \cos \theta_j  \Ket{b_j}\Ket{j} ,
\end{equation}
where 
\begin{align}
\ket{g_j} = \frac1{\sqrt{ L_j }} \sum_{ \{f(i,j) = 1 \}} \ket{i},
\end{align}
are the ``good'' data elements that are classified correctly by the $ j$th hyperplane and 
and 
\begin{align}
\ket{b_j } = \frac1{\sqrt{2^n- L_j }} \sum_{ \{f(i,j) = 0 \}} \ket{i}  
\end{align}
are the ``bad'' data elements that are not classified correctly.  
The angles are defined by 
\begin{align}
     \sin \theta_j & = \sqrt{\frac{L_j }{2^n}}  
     \label{anglerelation} \\
     \cos \theta_j & = \sqrt{\frac{2^n - L_j }{2^n}}   .
      \label{anglerelationcos}
\end{align}
Here we defined the number of data elements that are classified correctly by the $j$th hyperplane by
\begin{align}
L_j = \sum_{i=0}^{2^n-1} f(i,j) .  
\end{align}

Now we define the Grover operator 
\begin{align}
G = H^{\otimes n} S H^{\otimes n} U_f
\label{groveroperator}
\end{align}
where 
\begin{align}
    S := 2 |0 \rangle \langle 0 |^{\otimes n } - I
\end{align}
and the $ H $ are single qubit Hadamard gates.  We know that this operator acts as a rotation operation with angle $2\theta_j$ between the  ``good'' subspace  and the ``bad'' subspace, rotating $\Ket{g}\Ket{j}$ into $\Ket{b}\Ket{j}$ \cite{nielsen2002quantum}.  Hence, one application of the Grover operator rotates the state (\ref{eq:sincosstate}) to 
\begin{equation}
G \Ket{+}^{\otimes n}\Ket{j} = \sin 3\theta_j \Ket{g_j }\Ket{j} + \cos 3 \theta_j \Ket{b_j }\Ket{j}.
\end{equation}

To achieve our goal of version space training, we wish to find the hyperplane $ j $ such that $ f(i,j) = 1 $ for all $ i $.  In terms of the angle $ \theta_j $, according to  (\ref{anglerelation}), this means we require $\theta_j=\pi/2$. Hence, estimating $\theta_j $ will give us information regarding which of the $ j $ hyperplanes classify all the data correctly.  To perform this estimation, we use the quantum phase estimation algorithm.  We note that this is much like what is done in quantum counting \cite{Boyer1998}, where the number of solutions to a Grover problem is estimated.  This was also used in generalizations of Grover search \cite{Byrnes2018}.

The eigenvectors of $G$ are
\begin{align}
\Ket{\epsilon^\pm_j } =& \frac{\Ket{g_j} \pm\text{i}\Ket{b_j}}{\sqrt{2}}\Ket{j}
\end{align}
with eigenvalues $\text{e}^{ \pm\text{i}2\theta_j}$.
Now consider performing phase estimation on the initial state (\ref{eq:sincosstate}).
We rewrite this state as 
\begin{equation}
\ket{+}^{\otimes}\ket{j} =\frac1{\sqrt{2}}(\text{e}^{-\text{i}\theta_j}\Ket{\epsilon^+_j } +\text{e}^{+\text{i}\theta_j}\Ket{\epsilon^-_j }) | j \rangle .  
\end{equation}
If we apply phase estimation using the operator $G$, we obtain
\begin{equation}
\label{eq:stateafterphaesi}
\frac1{\sqrt{2}}(\text{e}^{-\text{i}\theta_j}\Ket{\epsilon^+_j} \ket{j} \ket{s^+} +\text{e}^{+\text{i}\theta_j}\Ket{\epsilon^-_j}  \ket{j} \ket{s^-}),
\end{equation}
where the output registers are $ | s^\pm \rangle  $ are the binary representations of the angle $ \theta_j $
\begin{align}
\frac{2\theta_j}{2\pi} & \approx  \sum_{n=1}^{l} \frac{s^+_n}{2^{n} },  \nonumber \\
\frac{2\pi-2\theta_j}{2\pi} & \approx \sum_{n=1}^{l} \frac{s^-_n}{2^{n}}, 
\label{approxangle}
\end{align}
where $ s^\pm_n $ denotes the $ n$th binary digit of $ s^\pm $.  This is only an approximate relation because the $ s^\pm $ are only accurate to $ l $ digits of binary precision.  In the limit $ l \rightarrow \infty $ the relations become equal.   The phase estimation algorithm can output $s^\pm $ with bounded error and $O(2^l)$ queries of $G$. A measurement of the register will collapse the state to one of the two terms in the superposition (\ref{eq:stateafterphaesi}) and  yield the value $ \pm \theta_j $ with probability $1/2$ for each case. In our case, no measurement is made. 

Let us now consider how many digits of precision $ l $ are required in the register to perform the phase estimation.  We wish to minimize $ l $ since a high precision of $\theta_j $ will require a larger oracle query count.
A key observation is that for our problem, we do not in fact need to know what $\theta_j $ is to full precision.  All that is necessary to construct $ g $ is to know whether $\theta_j=\pi/2$ or not.  
To achieve this, we must work out what is the minimum number of qubits needed such that we can distinguish between  $\theta_j=\pi/2$, and the closest case to this.  

Firstly, if $\theta_j=\pi/2$, we have $ s^\pm = 1/2$, which in binary form corresponds to 
\begin{align}
| s^\pm \rangle  = | 1 00 \dots 0 \rangle .  
\label{correctoutput}
\end{align}
The closest case to this is if there are all but one of the $ f(i,j) =1 $, such that $ L_j = 2^n - 1$.    In this case from Eq.~(\ref{anglerelationcos}) we have
\begin{align}
\cos \theta_j = \frac1{\sqrt{2^n}} .  
\end{align}
The angular difference of this from the $ g(j) =1 $ case is $ \delta = \pi/2 - \theta_j $, which is 
\begin{align}
\sin \delta \approx   \delta  = \frac1{\sqrt{2^n}} = \frac1{2^{n/2}} , 
\end{align}
where we used the small angle approximation.  We see that the minimum resolution that we must measure $ \theta_j $ with occurs with a bit precision of $n/2$. In other words, we only need to calculate  $\sim n/2$ bits of $s$ to distinguish the cases of $\theta_j=\pi/2$ and  $\theta_j < \pi/2$. 

We can now describe the full algorithm for perceptron training. The way to construct $ U_g $ is shown in Fig. \ref{fig:qcircuit_simand}. 
On completing the phase estimation algorithm, it is possible to know whether a given hyperplane $ j $ is in the version space or not, by checking whether the output register is in the state (\ref{correctoutput}).  
In order to produce the phase as given in (\ref{eq:U_g}),  a multi-qubit controlled-$Z$ gate according to the state (\ref{correctoutput}) is applied to the output register for phase estimation.  However, since the register is entangled with the data qubits (\ref{eq:stateafterphaesi}), and the Grover oracle needs to be called multiple times, we will need to uncompute the phase estimation algorithm such as to return to the starting state (\ref{eq:sincosstate}). 
After performing the uncompute step, this completes the $ U_g $ gate. From here, we can simply perform quantum search  with the oracle $ U_g $ to search through the $ K $ hyperplanes.  

Now let us estimate the oracular complexity of the circuit.  For a register with $ l =  n/2$ bits, $ O(\sqrt{N}) $ calls of $ U_f $ are necessary to run the phase estimation algorithm once. The uncompute step requires only a factor of 2 additional resources and does not change the complexity. The Grover step requires $ O(\sqrt{K}) $ calls, and hence we achieve our aim of performing perceptron training in $ O(\sqrt{NK}) $ steps.  

The result of phase estimation is in fact only correct with high probability, due to the approximate nature of  (\ref{approxangle}).  This means that the $U_g$ is imperfect, and can give the incorrect phase for some cases, although the error for this will be bounded.  To account for this, we use quantum search with bounded-error input instead of regular Grover search. In addition, we must also show that the multi-qubit controlled-$Z$ gate will give the same phase for the two terms in (\ref{eq:stateafterphaesi}).   We also show the optimality of the algorithm by relating it to a two-level AND-OR tree.  We give a more detailed argument for the construction of $ U_g $ in the next section.




\section{Quantum version space training: More detailed proof}
\label{sec:qversion_training}

In this section, we give a more detailed construction of  $U_g$ and provide the relevant proof.  We then show that  $U_g$ with bounded-error quantum search can solve the training problem. The optimality of our algorithm is also proven.

\begin{figure}[t]
    \centering
    \includegraphics[width=0.49\textwidth]{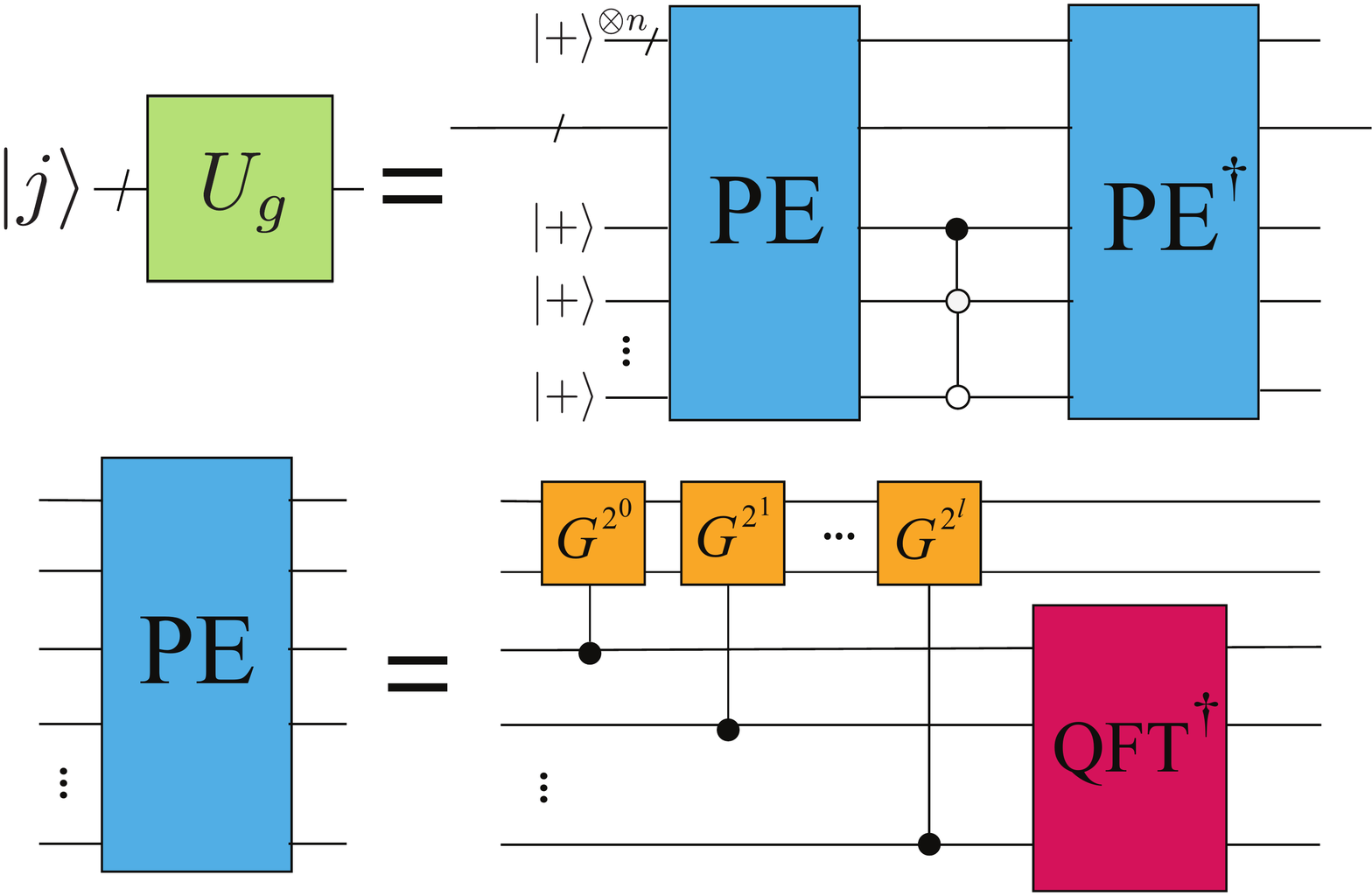}
    \caption{The quantum circuit for constructing the gate $U_g$. PE is the phase estimation circuit, $ G $ is the Grover operator (\ref{groveroperator}), and QFT$^\dagger$ is the inverse quantum Fourier transform.  }
    \label{fig:qcircuit_simand}
\end{figure}

\begin{algorithm}[t]
\begin{algorithmic}
\Require{
\Statex $n\in \mathbb{N}$
\Statex $k\in \mathbb{N}$
\Statex $j\in \set{0,1}^k$ 
\Statex $U_f$
\Comment{Quantum oracle in  \ref{eq:quantum_oracle}}
}
\Ensure{$\Ket{\psi}\in\mathcal{H}_2^{\otimes *}$}
\Function{SimAnd}{}
\State $l \gets \lceil n/2\rceil +3$ \Comment{number of bits for phase estimation}
\State $\Ket{\psi} \gets \Ket{+}^{\otimes n}\otimes\Ket{j}\otimes \Ket{+}^{\otimes l}$
\State  $S\gets 2\Ket{0}^{\otimes n}\bra{0}^{\otimes n}-I^{\otimes n}$
\State $G \gets \left((H^{\otimes n} S H^{\otimes n})\otimes I^k\right) U_f $
\State $\Ket{\psi} \gets \textsc{phaEst}$ $(G,l)\Ket{\psi} $
\If{$\underline{\Ket{\psi}[n+k+2]}=\underline{0} \wedge\cdots \wedge \underline{\Ket{\psi}[n+k+l]}=\underline{0} $ }
\State $Z[n+k+1]$ \Comment{Pauli $Z$ operator}
\EndIf
\State
$\Ket{\psi}\gets\textsc{phaEstInv}(O,l)\Ket{\psi}$
\State Return $\Ket{\psi}$
\EndFunction
\end{algorithmic}
\caption{Constructing gate $U_g$ with $O(\sqrt{N})$ queries to $U_f$.}
\label{al:simulating-AND}
\end{algorithm}

With access to $U_f$,  we can simplify Problem~\ref{pro:perceptron_version-space_training}
to the following search problem.
\begin{problem}[\textbf{Multi-criterion search}]
\label{problem:multi-criteria_search}
Given a Boolean function $f:[N]\times[K]\to\set{0,1}:(i,j)\to z$, where  $i\in [N],j\in[K], z\in \set{0,1}$, output a $j\in [K]$ such that $f(i,j)=1$ for all $ i\in[N]$ with bounded error or output~$-1$ if no such $j$ exists with bounded error.
\end{problem}
We call this problem a \textit{multi-criterion search} because $i$ can be viewed as a criterion.
By setting $N=1$, we have the regular Grover search, of which the search space is $[K]$ and one is asked to return a variable evaluated to be 1 for only one Boolean function. 
When $N\ge2$, the returned variable $j\in[K]$ in Problem~\ref{problem:multi-criteria_search} must evaluate to~$1$ for all functions $f_{i}(j)=f(i,j):[K]\to\set{0,1}$, i.e., there are multiple criteria.

One can also interpret this problem from the perspective of function inverse.
Unsorted search can be formalized as finding a variable that evaluates to~$1$ given the access to the oracle of the function.
The multi-criterion search can also be understood as a function inverse problem whereas the given oracle $U_f$ does not directly evaluate the desired function  $g(j)=\bigwedge_i f(i,j)$, so one needs to construct an oracle $U_g$ that evaluates the desired function $g$.

The implementation of $U_g$ in (\ref{eq:U_g}) with $\sqrt{N}$ calls of $U_f$ is shown as quantum circuit in Fig.~\ref{fig:qcircuit_simand} and 
the pseudocode of this algorithm is shown in Algorithm~\ref{al:simulating-AND}. 
Phase estimation and its inverse are denoted as  $\textsc{phaEst}$ and $\textsc{phaEstInv}$, respectively. 
%
\begin{lemma}
\label{lemma:sim_and}
For an input state $ \Ket{j}$, the output state of Algorithm~\ref{al:simulating-AND} is $ (-1)^{\tilde{g} (j)}\Ket{j}$ ignoring the ancillary qubits,
where the probability of obtaining $\tilde{g} (j)=g(j)$ is $ \ge 2/3 $ for all  $j\in\set{0,1}^k$.
\end{lemma}
\begin{proof}
The proof is given in Appendix \ref{prooflemma1}. In Appendix \ref{app:resources} we discuss the resources required for realizing a controlled-$ G $ gate and the qubit counts required to realize $ U_g $. 
\end{proof}

Lemma 1 shows that we can construct a gate $U_{\tilde{g}} $ satisfying $U_{\tilde{g}}  \Ket{j}=(-1)^{\tilde{g}(j) }\Ket{j}$ such that $\tilde{g} (j)=g(j)$ with high probability.
As the constructed $U_{\tilde{g}}  $ is not completely faithful,  we need to use the quantum search on bounded-error input \cite{Hoyer2003Quantum} instead of the regular Grover search
\begin{equation}
\textsc{bEqseArch}(k,U_{\tilde{g}}) \to j \in\set{0,1}^k~\text{or} -1 .     
\end{equation}
One can obtain an answer to Problem~\ref{problem:multi-criteria_search} and Problem~\ref{pro:perceptron_version-space_training} with $O(\sqrt{K})$ queries to $U_{\tilde{g}} $ thus $O(\sqrt{NK})$ queries of $U_f$.

To solve perceptron training via the version space method, one still needs to know how many hyperplanes need to be sampled.
It was proven in Ref.~\cite{kapoor2016quantum} that a random sampled hyperplane from a spherical Gaussian distribution where the mean is the zero vector and the covariance matrix is the identity matrix  perfectly classifies the given data set separated by margin~$\gamma$ with probability $\Theta(\gamma)$. Thus
$K$ scaling as $\nicefrac1{\gamma}$ makes sure that there is at least one sampled hyperplane that is in the version space with high probability.
We conclude that perceptron training can be solved with bounded error and query complexity $O\left(\sqrt{\nicefrac{N}{\gamma}}\right)$, achieving a quadratic speedup compared to the algorithm in Ref.~\cite{kapoor2016quantum}.

\section{Optimality}

\label{sec:optimality}

In this section, we prove the query complexity lower bound to solve Problem~\ref{problem:multi-criteria_search} is $\Omega\left(\sqrt{NK}\right)$; thus, our quantum training algorithm is optimal in the sense that the number of queries to $U_f$ can only be reduced up to a constant. 
We show that evaluating a two-level AND-OR tree can be  reduced to finding the solutions of Problem~\ref{problem:multi-criteria_search}.
As it is already proven the query complexity of evaluating a two-level AND-OR tree with $NK$ variables is $\Omega(\sqrt{NK})$, we have the query complexity of multi-criterion quantum search is also $\Omega(\sqrt{NK})$.
\begin{lemma}
\label{lemma:lowerbound}
The query complexity lower bound to solve multi-criterion search (Problem~\ref{problem:multi-criteria_search}) is $\Omega (\sqrt{NK})$.
\end{lemma}
\begin{proof}
First we introduce AND-OR trees, which correspond to  Boolean formulas only consisting of  $\operatorname{AND}$ and $\operatorname{OR}$ gates.  
For example, the following Boolean formula on $NK$ bits 
\begin{align}
\label{eq:AND-OR-bit}
&\operatorname{AND-OR}_2(z_0,\ldots,z_{NK-1})\\
=&\left(z_0\wedge\cdots\wedge z_{N-1}\right)\vee\cdots\vee
\left(z_{NK-N}\wedge\cdots\wedge z_{NK-1}\right) \nonumber
\end{align}
can be represented by a two level AND-OR tree with one OR gate acting on $K$ bits  and $K$ AND gates acting on~$N$ bits.
\begin{figure}[t]
    \centering
    \includegraphics[width=0.5\textwidth]{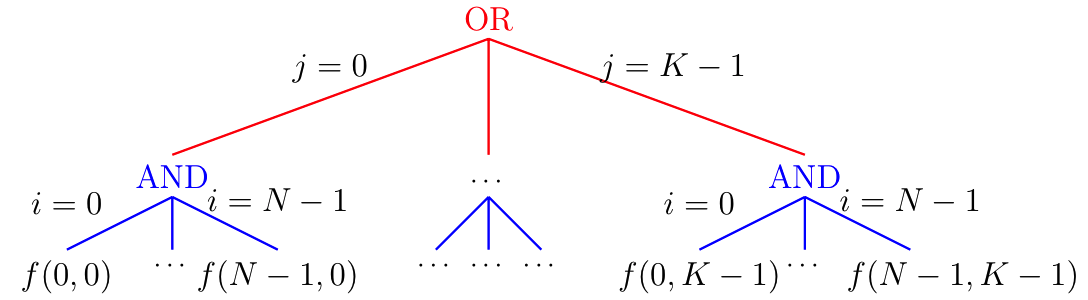}
    \caption{Two-level AND-OR tree that represents (\ref{eq:AND-OR}).}
    \label{fig:AND-OR_Tree}
\end{figure}
If we assume the bit $z$ can be accessed by an oracle $f(i,j)=z_{i+j N}$, in which $i\in[N], j\in[K]$, then the Boolean formula (\ref{eq:AND-OR-bit}) becomes 
\begin{equation}
\label{eq:AND-OR}
\begin{aligned}
\operatorname{AND-OR}(f) =
\bigvee_{j \in [K]}\left(\bigwedge_{i \in [N]} f\left(i,j\right)\right),
\end{aligned}
\end{equation}
as shown in Fig.~\ref{fig:AND-OR_Tree}.
The internal nodes of the tree are logical gates acting on its children, i.e., the nodes of the next level that are connected to them.

The reduction from evaluating (\ref{eq:AND-OR}) to a multi-criterion search is actually quite simple.
Suppose algorithm $\mathcal{A}$ solves the multi-criterion search problem.
If $\mathcal{A}$ outputs~$-1$, then there does not exist a column that is all 1s, so it is easy to see that $\operatorname{AND-OR}(f)= 0$;
otherwise, if $\mathcal{A}$ outputs any $j$, one obtains $\operatorname{AND-OR}(f)= 1$. 

It is proven in Ref.~\cite{ambainis2002quantum} that the query complexity to evaluate Eq.~(\ref{eq:AND-OR}) is $\Omega(\sqrt{NK})$, which completes the proof.
\end{proof}

We can formulate all above result as the following theorem:
\begin{theorem}
Given a data set $\Set{(\bm{x}_i,y_i)}_{i\in[N]}$ separated by margin~$\gamma$ and hyperplanes sampled from a sphere Gaussian distribution, the number of queries of $U_f$ needed to output a perceptron $\bm{p}$ that classifies all data correctly with probability greater than $2/3$ is $\Theta\left(\sqrt{\nicefrac{N}{\gamma}}\right)$.
\end{theorem}

\begin{proof}
It has been proven that the version space training of perceptron is equivalent to  multi-criterion search with access to $U_f$.
Combining Lemma~\ref{lemma:sim_and} and Lemma~\ref{lemma:lowerbound}, it is easy to see the complexity to solve multi-criterion search is $\Theta(\sqrt{NK})$, which completes the proof.
\end{proof}
In other words, since our algorithm as given in Sec.~\ref{sec:qversion_training} attains the lower bound scaling of $\Omega\left(\sqrt{NK}\right)$, we conclude that our proposed algorithm has an optimal scaling.

\section{Conclusion}
\label{sec:conclusion}
We have proposed a new quantum perceptron training algorithm that improves the query complexity from $O\left(N \sqrt{K}\right)$ to $\Omega\left(\sqrt{NK}\right)$.   We have shown how to construct an oracle $ U_g $ given an oracle $ U_f$, which provides the information of whether a given data point is classified by a given hyperplane.  This is achieved by using quantum counting, where the number of points that are classified correctly is counted.  The key point that results in the reduction of complexity results from the fact that for perceptron training, one only needs to distinguish the case where all the points are classified correctly, and all the remaining cases.  Determining the minimal number of qubits in phase estimation for quantum counting, one finds a quadratic speedup in comparison to the classical case. This results in a quadratic speedup in query complexity compared to previously best known proposed quantum training protocols.
Optimality of this procedure is found by reducing the version space training to a multi-criterion search, and showing the equivalence to a two-level AND-OR tree.  By showing the same complexity as the bound in Ref.~\cite{ambainis2002quantum}, this shows that our algorithm is optimal.  

As neural networks are constructed using a collection of perceptrons connected with each other,  our method  for training a single perceptron can be potentially generalized to train complex neural networks.  The reduction in scaling from $ N $ to $ \sqrt{N}$ is potentially extremely powerful since the size of the dataset is typically very large, and in this case a quadratic speedup is considerable.  Here we have not considered how to construct $U_f$ from the basic gates, so whether our advantage in query complexity can be transformed to the advantage in time complexity remains unknown, which is usually much harder to analyze~ \cite{cornelissen2020span,Belovs2012}.

\acknowledgments
T.\ B.\ is supported by the National Natural Science Foundation of China (62071301); State Council of the People’s Republic of China (D1210036A);
NSFC Research Fund for International Young Scientists (11850410426); NYU-ECNU Institute of Physics at NYU Shanghai; the Science and Technology Commission of Shanghai Municipality (19XD1423000);
the China Science and Technology Exchange Center (NGA-16-001); the NYU Shanghai Boost Fund. 
B.\ C.\ S.\ has support from Natural Science and Engineering Research Council of Canada (NSERC) and from the National Natural Science Foundation of China~(NSFC) with Grant No.~11675164.

\appendix

\section{Proof of Lemma \ref{lemma:sim_and}}
\label{prooflemma1}

Here we present a proof for Lemma~\ref{lemma:sim_and}.
The quantum registers consisting of $n$, $k$, and $l$ qubits are referred to as the first, second, and third quantum registers, respectively.

From Eq.~(\ref{eq:stateafterphaesi}), we found that  the quantum state after performing phase estimation 
$\textsc{phaEst} \left(G,l\right)$ on the state $ \Ket{+}^{\otimes n} \Ket{j}  \Ket{+}^{\otimes l} $ is 
\begin{equation}
\frac1{\sqrt{2}} \left(
\text{e}^{-\text{i}\theta_j}\Ket{\epsilon_j^+}\Ket{j}\Ket{s^+}+\text{e}^{+\text{i}\theta_j}\Ket{\epsilon_j^-} \Ket{j}\Ket{s^-}
\right).    
\end{equation}
The multi-qubit controlled $ Z $ gate adds a phase $-1 $ based on the state in the third quantum register
\begin{align}
\label{eq:phase_s+-}
\frac1{\sqrt{2}}(
(-1)^{p(s^+)} \text{e}^{-\text{i}\theta_j}\Ket{\epsilon_j^+}\Ket{j} \Ket{s^+}+\nonumber\\
(-1)^{p(s^-)}\text{e}^{+\text{i}\theta_j}\Ket{\epsilon_j^-} \Ket{j} \Ket{s^-}),    
\end{align}
where
\begin{equation}
    p(s)= s_1\wedge \left(\wedge_{i=2}^{l}(\neg s_i)\right)\in\set{0,1}
\end{equation}
and $\neg$ is the Boolean operator NOT acting on bit.

If  $p(s^+)=p(s^-)$, the added phase will not affect the relative phase of the quantum state.
After applying the inverse of phase estimation, state (\ref{eq:phase_s+-}) is transformed to $(-1)^{p(s^\pm)}\Ket{+}^{\otimes n}\Ket{j}\Ket{+}^{\otimes l}$.
Ignoring the ancillary qubits in the first and the third quantum register, we get $(-1)^{p(s^\pm)}\Ket{j}$. In order to prove this procedure implements $U_g$,
 we need now to prove that the added phase should faithfully correspond to the value of $g(j)$, i.e.,  $p(s^+)=p(s^-)=g(j)$ holds for every $j\in\set{0,1}^k.$

First, consider the case when $g(j)=1$, i.e., $U_f\Ket{i,j}=-\Ket{i,j}, \forall i\in \set{0,1}^n$.
In this case, it is easy to see the angle in in (\ref{eq:sincosstate}) is $\theta_j= \pi/2$, so $2\theta_j=2\pi-2\theta_j=\pi$ and  the output bitstrings satisfy $s^+=s^-$.
The output bitstring in phase estimation also satisfies \begin{equation}
    \frac{2\theta_j}{2\pi} =\frac1{2} =\sum_{i=1}^{\infty} s^+_i2^{-\text{i}},
\end{equation}
so we have $s^+_1=s^-_1=1$ and $s^+_i=s^-_i=0, \forall i\ge 2$.
It is then easy to obtain $p(s^+)=p(s^-)=1$. 

When $g(j)=0$, we have $\theta_j< \pi/2$, so
$\frac{2\theta_j}{2\pi} =\sum_{i=1}^{\infty} s^+_i2^{-\text{i}}<\frac1{2}$, so the first bit of $s^+$ is $s^+_1=0$, leading to $p(s^+)=0$.
As $2\pi-2\theta_j >\pi$, we have $s^-_1=1$, so the value of $p(s^-)$ depends on whether $\exists~2\le i\le l, s^-_i=1$ holds.  
Next, we prove that when $l=\lceil n/2\rceil+3$, then there exists $2\le i\le l$ such that $s^-_i=1$.

Consider the case where $\sin^2 \theta_j= \frac{2^n-m}{2^n} = 1 - \cos^2\theta_j$ and $1\le m\le 2^n$,  we have $\theta_j = \arccos{\sqrt{\frac{m} {2^n}}}$ and
\begin{align}
\frac{\theta_j}{2\pi} = & \frac1{4} - \frac1{2\pi} \sqrt{\frac{m} {2^n}}-O(\sqrt{\frac{m^3} {2^{3n}}}),    \\
\frac{2\pi-2\theta_j}{2\pi} 
= &\frac1{2} + \frac1{2\pi} \sqrt{\frac{m} {2^n}}+O(\sqrt{\frac{m^3} {2^{3n}}}) \nonumber\\
\label{eq:sum_s_i/2^i_-2theta}
= &\frac{s_1^-}{2}+\cdots+\frac{s_i^-}{2^i}+\cdots \ge \frac1{2}+\frac1{2^{\lceil n/2\rceil + 3}}
\end{align}
From Eq.~(\ref{eq:sum_s_i/2^i_-2theta}) and the fact that $s_1^-=1$, one can easily see that there exists $s_i^-=1 (2\le i\le \lceil n/2\rceil+3)$ so $p(s^-)=0$ holds as well.
In summary, $p(s^+)=p(s^-)=g(j)$ is true.

\section{Resources counts of the oracle}
\label{app:resources} 

Note that in Algorithm~\ref{al:simulating-AND}, a controlled-$G$ gate is needed, which implicitly means we need to apply controlled-$U_f$ instead of $U_f$ itself.
One may wonder whether the controlled-$U_f$ is an equivalent resource to $U_f$ and if it is a fair comparison between how many times the controlled-$U_f$ and  $U_f$ itself are called.
We show here that both $U_f$ and controlled-$U_f$ can be constructed by calling to $U_f^\prime$ different by a factor of 2, where
\begin{equation}
U_f^\prime \Ket{i,j,a} = \Ket{i,j,a+f(i,j)},    
\end{equation}
so they are equivalent resource up to a constant factor in terms of $U_f^\prime$.
It is commonly known by setting the last qubit to state $\Ket{-}:=\frac{\Ket{0}-\Ket{1}}{\sqrt{2}}$, one has
\begin{equation}
 U_f^\prime \Ket{i,j}\Ket{-} = (-1)^{f(i,j)}\Ket{i,j}\Ket{-},
\end{equation}
which achieves $U_f$ with one ancillary qubits.
On the other hand, with a controlled-$Z$ (CZ) gate acting on the first qubit (control qubit) and the last qubit (target qubit) and calling to $U_f^\prime$ twice, we have
\begin{align}
 &U_f^\prime\operatorname{CZ} U_f^\prime\Ket{c,i,j,0} = U_f^\prime \operatorname{CZ}\Ket{c,i,j,f(i,j)} \\
 =&U_f^\prime (-1)^{c  f(i,j)}\Ket{c,i,j,f(i,j)} =  (-1)^{c  f(i,j)}\Ket{c,i,j,0}, \nonumber
\end{align}
which is the controlled-$U_f$ using one ancillary qubit.
Therefore, the controlled-$U_f$ and $U_f$ require equivalent resources up to constant factor.  Hence, this does not undermine our quadratic speedup.  

In addition, it is also easy to see from Fig.~\ref{fig:qcircuit_simand} that the space cost to construct $U_g$ is about $3n/2 +k$ qubits in our proposal while the space cost in Ref.~\cite{kapoor2016quantum} is only $n+k$. An extra $l=\lceil n/2\rceil +3 $ qubits are needed but the asymptotic scaling of space resource required in both cases are in $O\left(\log_2(NK)\right)$.

\bibliographystyle{apsrev}
\bibliography{paperrefs}
\end{document}